\documentclass[12pt]{article}
\usepackage{subcaption}
\usepackage{graphicx}
\usepackage{authblk}
\usepackage{amsmath}
\usepackage{siunitx}
\usepackage{bm}
\usepackage{upgreek}
\usepackage[sort]{scicite}
\usepackage{color}
\usepackage{ulem}

\newenvironment{sciabstract}{%
\begin{quote} \bf}
{\end{quote}}

\makeatletter
\renewcommand{\@seccntformat}[1]{}
\makeatother

\makeatletter
\def\p@subsection{}
\makeatother

\usepackage{xr}
\definecolor{revclr}{RGB}{255,0,0} 

\title{Broadband quadrature-squeezed vacuum and nonclassical photon number correlations from a nanophotonic device}

\author{V.D. Vaidya,$^1$ B. Morrison,$^1$ L.G. Helt,$^1$ R. Shahrokshahi,$^1$ D.H. Mahler,$^1$ M.J. Collins,$^1$ K. Tan,$^1$ J. Lavoie,$^1$ A. Repingon,$^1$ M. Menotti,$^1$ N. Quesada,$^1$ R.C. Pooser,$^2$ A.E. Lita,$^3$ T. Gerrits,$^3$ S.W. Nam,$^3$ Z. Vernon,$^1$
\\
\normalsize{$^1$ Xanadu, Toronto, ON, M5G 2C8, Canada}\\
\normalsize{$^2$ Oak Ridge National Laboratory, Oak Ridge, Tennessee 37831, USA} \\
\normalsize{$^3$ National Institute of Standards and Technology (NIST), 325 Broadway, Boulder, Colorado 80305, USA}
}

\date{}

\begin{document}

\baselineskip24pt

\maketitle 

\begin{sciabstract}
We report demonstrations of both quadrature squeezed vacuum and photon number difference squeezing generated in an integrated nanophotonic device. Squeezed light is generated via strongly driven spontaneous four-wave mixing below threshold in silicon nitride microring resonators. The generated light is characterized with both homodyne detection and direct measurements of photon statistics using photon number-resolving transition edge sensors. We measure $1.0(1)$~dB of broadband quadrature squeezing (${\sim}4$~dB inferred on-chip) and $1.5(3)$~dB of photon number difference squeezing (${\sim}7$~dB inferred on-chip). Nearly-single temporal mode operation is achieved, with measured raw unheralded second-order correlations $g^{(2)}$ as high as $1.95(1)$. Multi-photon events of over 10 photons are directly detected with rates exceeding any previous quantum optical demonstration using integrated nanophotonics. These results will have an enabling impact on scaling continuous variable quantum technology. 
\end{sciabstract}

\bigskip

\section{INTRODUCTION}
Squeezed light is an essential resource for quantum optical information processing. Continuous variable (CV) photonic architectures for quantum computation and simulation \cite{weedbrook2012gaussian} demand high quality scalable devices for generating squeezed light, which is the fundamental building block for many photonic quantum information processing applications. These include CV quantum computation \cite{braunstein2005quantum} and Gaussian boson sampling \cite{hamilton2017gaussian}. The latter is one of the most promising avenues to the achievement of near-term quantum advantage, and accommodates a host of intriguing use cases, including molecular vibronic spectrum simulations \cite{huh2015boson}, and quantum embeddings of graph problems such as the identification of dense subgraphs \cite{arrazola2018using}, perfect matchings \cite{bradler2017gaussian}, graph isomorphism \cite{bradler2018graph}, and graph similarity \cite{schuld2020measuring}. Gaussian boson sampling can also be used to predict molecular docking configurations \cite{banchi2019molecular}. A scalable source of squeezed light is needed to implement all of these, and could also be used to enhance optical sensing near the quantum limit \cite{caves1981quantum}. A natural platform to explore for such scalable squeezed light sources is integrated photonics: the stability and repeatable manufacturability offered by modern lithographic techniques hold much promise for realizing useful quantum technologies at scale. The integration of quantum light \textit{sources} is especially important in variants of quantum computation based on continuous variables. This is due to the acute phase sensitivity of squeezed states, which serve as the basic resource for nonclassicality in such variants. 

Progress to date on chip-integrated squeezing has been limited. Lenzini \textit{et al.} \cite{lenzini2018integrated}, Mondain \textit{et al.} \cite{mondain2019chip}, and Stefszky \textit{et al.} \cite{stefszky2017waveguide} have demonstrated devices based on low index contrast lithium niobate waveguides with large cross-sections, using periodically poled waveguide segments for squeezed light generation, in combination with linear optics on a monolithic chip.  While this platform is attractive for its high second-order nonlinearity and electro-optic response, as well as its ease of integration with fiber components, currently it lacks the scalability offered by nanophotonic systems: the low transverse confinement prohibits the monolithic integration of high-depth circuits with hundreds of optical elements. Dutt \textit{et al.} \cite{dutt2015chip} made the first forays into high index contrast nanophotonic systems for squeezing by demonstrating intensity difference squeezing from a silicon nitride microring optical parametric oscillator (OPO) driven above threshold. However, quadrature squeezing was not achieved, and the demonstrated bright ``twin beam" type squeezing above threshold is necessarily accompanied by large classical mean fields and very high levels of excess noise, rendering it incompatible with experiments or architectures that require photon counting. 

Among such architectures, quantum sampling (which includes the applications of Gaussian boson sampling \cite{huh2015boson,arrazola2018using,bradler2017gaussian,schuld2020measuring,banchi2019molecular,bradler2018graph}) places even more stringent demands on the nature of the squeezed light source employed: in addition to being scalable, such a source must be capable of producing squeezed states in a single temporal mode \cite{vernon2019scalable}. This requirement is often overlooked, since it can often be avoided in single photon based quantum protocols by adequate spectral filtering, or by post-selecting on detection events that occur within a common, narrow temporal window. Unfortunately, such strategies cannot be used for CV quantum sampling: spectral filtering imposes unacceptable losses on the system, the effects of which cannot be circumvented by post-selecting on successful detection events; and temporal windowing cannot be applied on detection events in which many-photon contributions are significant, owing to the low timing resolution of available photon-number resolving systems \cite{rosenberg2005noise}. Since many-photon events are in general important throughout the application space of CV quantum sampling, and losses have significant deleterious effects on both fidelities and count rates \cite{qi2020regimes}, it becomes vital to engineer a source which itself natively produces squeezed states in a single temporal mode. This feature has proven challenging across all platforms, and has been most thoroughly investigated in periodically poled crystalline $\chi_2$ waveguides \cite{eckstein2011highly,harder2013optimized,harder2016single} wherein approaching single temporal mode operation requires very careful dispersion engineering. These considerations make the temporal mode structure a key metric to assess when qualifying squeezed light sources for practical use.  

Here we report the demonstration of quadrature squeezed vacuum and photon number difference squeezing generated with an integrated nanophotonic device. We employ spontaneous four-wave mixing (SFWM) in silicon nitride microring resonators \cite{moss2013new}, a readily accessible and mature technology available on commercial fabrication platforms.  We measure the variance of all quadratures of the squeezed modes with balanced homodyne detection, and we also verify the compatibility of this source with photon counting experiments and sampling applications by directly measuring the photon statistics of the output using photon number-resolving transition edge sensors (TES) \cite{rosenberg2005noise}. Raw photon number correlations are assessed, and close attention is paid to the temporal mode structure of the generated light: unheralded second-order correlation measurements are performed to ensure our device can approach single temporal mode operation.  \cite{christ2011probing}.

\begin{figure}
    \centering
    \includegraphics[width=0.7\textwidth]{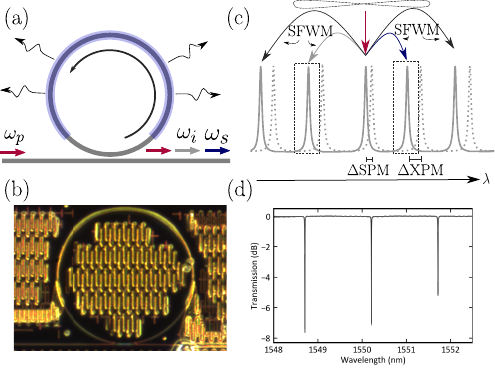}
    \caption{(a) Schematic of microring device showing resonator, side channel, microheater (blue), and scattering modes. (b) Optical microscope image of device. (c) Illustration of intraresonator spontaneous four-wave mixing process, showing frequency shifts $\Delta\mathrm{SPM}$ and $\Delta\mathrm{XPM}$ associated with self- and cross-phase modulation. (d) Representative transmission spectrum of microring device, showing three over-coupled resonances near 1550 nm. }
    \label{fig:scheme}
\end{figure}
The device itself is simple: both quadrature and photon number difference squeezing are generated in silicon nitride microring resonators point coupled to channel waveguides, with microheaters overlaid for resonance wavelength tuning and stabilization. Fabrication was carried out by a commercial foundry (Ligentec SA). The waveguide geometry consists of a stoichiometric $\mathrm{Si}_3\mathrm{N}_4$ wire fully clad in $\mathrm{SiO}_2$. The cross-sections used were $800$~nm x $1000$~nm for quadrature squeezing, and $800$~nm x $1650$~nm for photon number correlation measurements. This platform was selected for its low linear propagation loss, lack of two-photon absorption, and high third-order nonlinear parameter of approximately $1$~(Wm)$^{-1}$. To generate squeezing, SFWM was employed: a single strong pump (but weak enough to stay below any OPO thresholds), resonant with one of the ring resonances, generates multi-mode squeezed vacuum in a comb of neighbouring resonances (Fig. \ref{fig:scheme}). The nonlinear Hamiltonian $H_\mathrm{NL}$ describing the interaction between the pump mode and a particular pair of signal and idler resonances is given by \cite{vernon2015spontaneous}
\begin{align}\label{eqn:hamiltonian}
    H_\mathrm{NL} &= -\hbar\Lambda \bigg ( (b_\mathrm{P} b_\mathrm{P} b_\mathrm{S}^\dagger b_\mathrm{I}^\dagger + \mathrm{H.c} )  \nonumber \\ 
    &\quad\quad+ \frac{1}{2}b_\mathrm{P}^\dagger b_\mathrm{P}^\dagger b_\mathrm{P} b_\mathrm{P} + 2b_\mathrm{P}^\dagger b_\mathrm{P} (b_\mathrm{S}^\dagger b_\mathrm{S} + b_\mathrm{I}^\dagger b_\mathrm{I})\bigg ),
\end{align}
where $b_x$ is the annihilation operator associated with the resonator mode $x \in \lbrace \mathrm{P,S,I}\rbrace$, representing the pump, signal, and idler modes, respectively. Here $\Lambda$ is a coefficient representing the nonlinear coupling strength between the modes, and is well approximated by $\Lambda \approx \hbar \overline{\omega}v_g^2\gamma_\mathrm{NL}/L$, where $\overline{\omega}=(\omega_\mathrm{P}^2\omega_\mathrm{S}\omega_\mathrm{I})^{1/4}$, $v_g$ is the group velocity, $\gamma_\mathrm{NL}$ the waveguide nonlinear parameter, and $L$ the resonator round trip length. 

The first term in Eq. \ref{eqn:hamiltonian} leads to SFWM, in which two pump photons are annihilated and a pair of signal and idler photons are generated; taken to higher order, this process generates a squeezed state involving the signal and idler \cite{lvovsky2015squeezed,vernon2019scalable}. For a strong coherent and continuous wave (CW) input pump field we may treat the intraresonator pump classically, and replace $b_\mathrm{P}$ by its classical expectation value $\beta_\mathrm{P}=\overline{\beta}_\mathrm{P}e^{-i(\omega_\mathrm{P}+\Delta_\mathrm{P})t}$, where $\overline{\beta}_\mathrm{P}$ is a constant and $\Delta_\mathrm{P}$ the input pump field detuning from resonance. The effects of self- and cross-phase modulation (SPM and XPM, respectively), represented by the last two terms in Eq. (\ref{eqn:hamiltonian}), then manifest as pump power-dependent frequency shifts of the pump, signal, and idler resonances (Fig. \ref{fig:scheme}(c)). Note that in Eq. (\ref{eqn:hamiltonian}) we neglect the effects of SPM on the signal and idler modes, since these effects are negligible in our regime of interest well below the OPO threshold. Since the waveguides used in our devices for quadrature squeezing operate in the normal dispersion regime, these frequency shifts are detrimental for a single pump configuration, as they lead to an increased energy mismatch of the resonance frequencies away from the ideal operational point for SFWM of $2\omega_\mathrm{P}=\omega_\mathrm{S}+\omega_\mathrm{I}$. Despite this, significant quadrature squeezing at modest input power levels was still achieved without dispersion engineering.

The statistics of the output signal and idler fields can be calculated using cavity input-output theory; full details are available in the Supplementary Material. In the subsequent sections we report the measured quadrature and photon number statistics from the device, and then compare our results with the predictions based on this theory. 

\bigskip

\section{RESULTS}
\bigskip
\subsection{Quadrature squeezing}\label{sec:quadrature_squeezing}
\begin{figure*}[h!]
\centering
\begin{subfigure}{0.48\textwidth}
\includegraphics[width=1.0\textwidth]{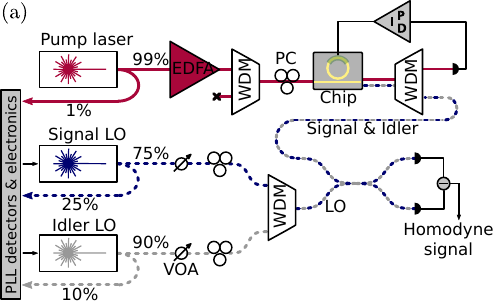}
\end{subfigure}
\begin{subfigure}{0.48\textwidth}
\includegraphics[width=1.0\textwidth]{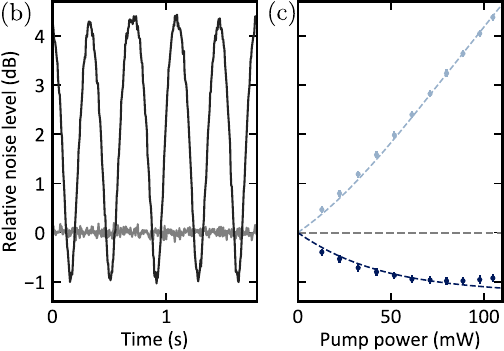}
\end{subfigure}
\caption{Quadrature squeezing: (a) Overview of experimental setup. Details in text and Supplementary Material. (b) Quadrature variance (black line) relative to shot noise (grey line) as a function of time while the local oscillator phase is ramped, exhibiting $1.0(1)$~dB of squeezing. Traces are obtained from the homodyne detector photocurrent fluctuations monitored on an electrical spectrum analyzer in zero-span mode at $20$ MHz sideband frequency, with resolution bandwidth 1 MHz, video bandwidth 300 Hz. (c) Maximum and minimum quadrature variances as a function of pump power for the $20$~MHz sideband, showing the power scaling of the squeezed and anti-squeezed quadratures. The upper and lower dashed lines are obtained by fitting to Eq. (\ref{eqn:variances}); the shot noise level is shown (dashed line at $0$~db).}\label{fig:quadrature_fig_exp}
\end{figure*}

Since the SFWM interaction is of the form $b_\mathrm{S}^\dagger b_\mathrm{I}^\dagger$, the quantum state of the resonator output can (in the absence of non-parametric effects that could contribute noise) be characterized as a two-mode squeezed vacuum state subjected to loss; this loss arises both from the imperfect escape efficiency of the cavity, as well as ``downstream" losses that are incurred at the chip outcoupling point, post-chip fiber components, and those due to detector inefficiency. This state can also be understood as a product of two \textit{single}-mode squeezed states, each having frequency support at both the signal and idler resonances, i.e., in modes described by annihilation operators $\widetilde{b}_\pm=(b_\mathrm{S} \pm b_\mathrm{I})/\sqrt{2}$. In what follows we adopt this perspective, and study the quadrature statistics of these composite \textit{bichromatic} modes \cite{marino2007bichromatic,embrey2016bichromatic}.

Normalizing the quadrature variance of the vacuum state to unity, the maximum and minimum quadrature variances $V_\mathrm{\pm}$ associated with a bichromatic mode in the output channel of the resonator having equal support at the signal and idler frequencies can be expressed as
\begin{eqnarray}\label{eqn:variances}
V_\mathrm{\pm}=1 + 4\eta g\left(2g\pm\sqrt{1+4g^2}\right),
\end{eqnarray}
where $\eta$ is the net collection efficiency (including the ring escape efficiency and all collection and detection losses), and $g$ is a dimensionless gain parameter defined as $g=\Lambda |\overline{\beta}_\mathrm{P}|^2/\overline{\Gamma}$, with $\overline{\Gamma}$ the dissipation rate of the signal and idler cavity modes (assumed for simplicity to be equal). This dissipation rate can be estimated from the full loaded quality factor $Q$ associated with a resonance at frequency $\omega$ via $\overline{\Gamma}\approx \omega/(2Q)$. Note that here we assumed the pump frequency is adjusted at each fixed input power to follow the pump resonance as its frequency shifts due to SPM and thermal effects. Doing so avoids crossing the OPO threshold at any power, and maintains the linear scaling of intraresonator circulating power with pump input power \cite{vernon2015strongly}.

\begin{figure}[h!]
\centering
\includegraphics[width=0.7\textwidth]{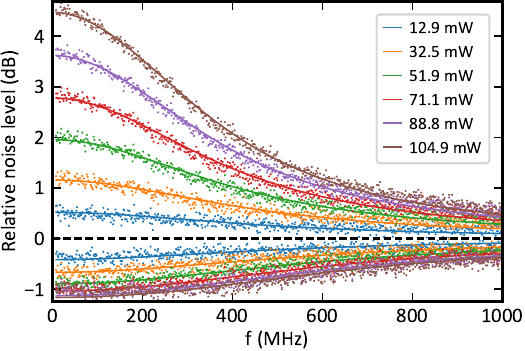}
\caption{Squeezing and anti-squeezing frequency spectrum from 20 MHz to 1 GHz at different pump power levels. Powers listed are inferred values on-chip in the input waveguide. Dashed line is shot noise level; Solid lines exhibit fits to the theoretical model (details in Supplementary Material), showing strong agreement with the measured data. }\label{fig:sqz_spec}
\end{figure}

To measure the quadrature variances $V_\pm$ of the modes of interest, balanced homodyne detection was performed on the device output. The experimental setup is illustrated in Fig. \ref{fig:quadrature_fig_exp}(a): a CW pump is coupled into the chip (and light extracted) via low-loss edge couplers. In the chip the pump excites a single resonance of the microring and generates squeezed light across multiple signal and idler pairs. One such pair of signal and idler modes is selected by off-chip wavelength filters for analysis, while the remaining pump light after the chip is monitored and used for feedback to stabilize the microring resonance frequency. Since the signal and idler resonances are separated in frequency from the pump by ${\sim}190$~GHz, beyond the capabilities of photoreceiver electronics, it was necessary to generate a bichromatic local oscillator \cite{embrey2016bichromatic} with frequency support at both $\omega_\mathrm{S}$ and $\omega_\mathrm{I}$, coherent and phase stable with respect to the pump. To do so, two separate CW lasers were tuned to the signal and idler frequencies and phase locked to each other, and to the pump, by beating them independently against the teeth of an optical frequency comb derived from a portion of the pump using a fast electro-optic modulator. This produced a bichromatic local oscillator that is phase stable relative to the pump, with approximately $4^\circ$ RMS phase noise between the pump and each local oscillator wavelength. The bichromatic local oscillator was then combined with the signal and idler light on a tunable fiber beam splitter set to $50/50$  splitting ratio, with the outputs incident on a commercial fast balanced amplified differential photoreceiver with $1$~GHz bandwidth. The difference photocurrent fluctuations were monitored on an electrical spectrum analyzer. As the local oscillator phase is ramped, the variance of different quadratures (Fig. \ref{fig:quadrature_fig_exp}(b)) can be identified relative to the shot noise level, which is measured by disconnecting the chip output.  This process was repeated for a range of input pump powers, yielding the curves presented in Fig. \ref{fig:quadrature_fig_exp}(c). Approximately $1.0(1)$~dB of squeezing is observed at the maximum input power at a sideband frequency of $20$~MHz. Considering the estimated collection and detection losses, and assuming that there is indeed no appreciable excess noise, we estimate that about $4$~dB of squeezing is available at the microring output on-chip; though this inference is affected significantly by the uncertainty in the loss estimate. The squeezing is limited both by available pump power, and by the finite escape efficiency of the resonator (${\sim}75$\%). We also note that the squeezing is broadband, limited by the resonance linewidths; appreciable squeezing was observed at sideband frequencies as high as $1$ GHz (limited by the detector bandwidth), as shown on the squeezing spectra illustrated in Fig. \ref{fig:sqz_spec}.

Apart from losses and limited pump power, another factor that can limit squeezing is the presence of excess noise added by processes other than parametric fluorescence. This is especially a concern for squeezing generated by third-order nonlinear optical interactions, in which the pump frequency is much closer to the measurement bands than is the case with more conventional downconversion-based squeezers. Indeed, our initial attempts to measure squeezing from similar devices within the pump resonance itself using fully degenerate four-wave mixing,  based on the proposal of Hoff \textit{et al.} \cite{hoff2015integrated}, were hampered by strong excess noise contributions even at sideband frequencies exceeding $500$~MHz.  We attribute this to thermorefractive fluctuations: though such contributions are usually only analyzed within a few MHz of the pump frequency, for the pump powers required to generate squeezing, thermorefractive fluctuations can scatter a non-negligible amount of light to much higher frequency sidebands. Recent work \cite{le2018impact, huang2019thermo} suggests that such noise can be measurable even at THz offsets from the pump. This attribution is corroborated by the results of a recent experiment carried out by Cernansky and Politi \cite{cernansky2019nanophotonic}, who attempted the Hoff \textit{et al.} scheme in a Sagnac loop configuration to suppress the bright pump.  While a small amount of quadrature squeezing at high sideband frequencies was reported, most of the mode was contaminated by excess noise that eliminated squeezing on all quadratures.  Aside from thermorefractive contributions, spontaneous Raman scattering (either in the microring device, channel waveguide, or fiber components used in the experiment) could also be present. Considering this ``mine field" of non-parametric effects that can corrupt squeezing in devices based on third-order nonlinearities, it is important to assess the presence of excess noise in our system generated inside the squeezing band. 

To gauge whether excess noise processes contribute to the measured variance, as shown in Fig. \ref{fig:quadrature_fig_exp}(c) we fit the measured behaviour of the variances $V_\pm$ to a theoretical model which assumes \textit{only} SFWM and loss, and no non-parametric sources of excess noise. The same is done for the squeezing spectra exhibited in Fig. \ref{fig:sqz_spec}. From the agreement of the fits with the data, we conclude that, within our measurement precision and at the pump power levels used, a model with no excess noise is adequate to account for the measured quadrature variances; our results can be explained very well by a squeezed vacuum state subjected to loss. Furthermore, the degree of loss inferred from the data is consistent with estimates based on direct loss measurements. We note that at the highest powers used, a very slight drop in the squeezing was observed; however, this is expected from the rapid growth of the noise in the anti-squeezed quadrature, which at finite phase precision negatively impacts the observed squeezing levels. More study is needed with improved phase precision to better assess this effect.

\bigskip
\subsection{Photon number correlations}\label{sec:photon_number_correlations}

\begin{figure}[h!]
\centering
\begin{subfigure}{0.7\textwidth}
\includegraphics[width=1.0\textwidth]{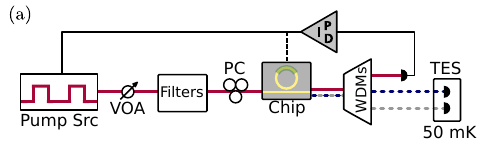}
\end{subfigure}
\begin{subfigure}{0.7\textwidth}
\includegraphics[width=1.0\textwidth]{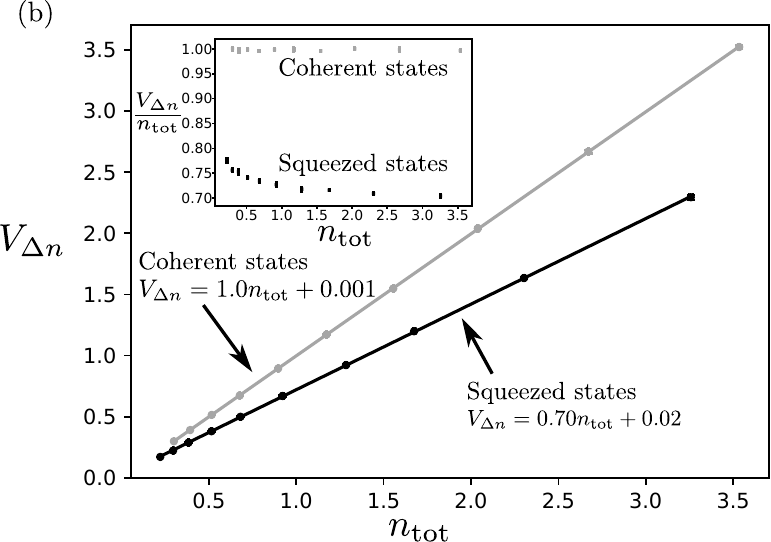}
\end{subfigure}
\caption{Photon number difference squeezing: (a) Overview of experimental setup. Details in text and Supplementary Material. (b) Measured photon number difference variance $V_{\Delta n}$ as a function of mean photon number $n_\mathrm{tot}$, obtained by varying pump power, for coherent states (grey) and squeezed states (blue) with linear fits (solid lines). The reduced slope for the squeezed state represents photon number difference squeezing. Inset: ratio between number difference variance and mean photon number as a function of mean photon number for coherent states (grey) and squeezed states (blue).}\label{fig:tes_fig_exp}
\end{figure}

While homodyne measurements are vital to assess quadrature squeezing, for many applications it is also important to verify the compatibility of a squeezed light source with photon counting. Homodyne detection provides exquisite mode selectivity of the probed quantum state, whereas most photon counting schemes are limited in their ability to extract information from only a single well-defined field mode without incurring unacceptably high losses. For example, at the powers required to generate squeezing, one might be concerned that \textit{broadband} spontaneous Raman scattering of the bright pump \cite{samara2019high} (either in the chip or in the various fiber components through which the pump must propagate) might place unreasonable spectral filtering requirements on the signal and idler paths, adding losses that would destroy quantum correlations. Additionally, as discussed in the Introduction, any multi-modedness of the light measured by a photon counting detection system will alter the measured statistics and corrupt the corresponding performance. An affirmative statement regarding the compatibility of a squeezed light source with photon counting therefore relates both to the greater experimental context as well as to the dynamics of the process that leads to squeezing. 

To verify the compatibility of our squeezed light source with photon counting, we performed photon number-resolving detection on the device output. For this experiment a microring resonator with a wider cross-section of 800 nm x 1650 nm was used, which resulted in a higher loaded quality factor of approximately $8\times 10^5$ for the three resonances (escape efficiency $80\%$). The device used for the quadrature squeezing experiment was selected to emphasize the broadband squeezing that can be generated in such microring devices; larger resonance linewidths (and thus lower quality factors) are required for that demonstration. While some photon number squeezing could be observed in the device used for quadrature squeezing, which displayed lower quality factors, results were much improved for the higher quality factor device. This is due to the higher sensitivity of photon number squeezing to the effects of broadband Raman noise generated in fiber components: our photon detectors no longer benefit from the extremely narrowband mode selectivity of homodyne detection, and instead must accept all the noise present within the 100 GHz passband of the filtering system employed. Higher quality factors permit weaker pumps to be used, thereby increasing the overall signal to noise ratio. 

As illustrated in Fig. \ref{fig:tes_fig_exp}(a), the generated signal and idler were first separated and filtered by wavelength division multiplexing components (WDMs) yielding a total extinction of above $120$~dB at the pump wavelength, while incurring less than $2$ dB of loss on the signal and idler. The output was then coupled to superconducting transition edge sensors \cite{rosenberg2005noise}, which provide photon number resolution up to about $10$ photons per channel. The number of signal and idler photons detected in each pulse window was recorded for a range of pump powers, and the photon statistics analyzed. 

For a lossy parametric fluorescence process without excess noise we expect the variance of the photon number difference $V_{\Delta n}$ per pulse between the signal and idler to satisfy
\begin{eqnarray}\label{eqn:photon_variance}
V_{\Delta n}= (1 - \eta)n_\mathrm{tot},
\end{eqnarray}
where $\eta$ is the total collection efficiency and $n_\mathrm{tot}=\langle n_\mathrm{S} + n_\mathrm{I}\rangle$ the total average photon number detected in the signal and idler. For a coherent state, this variance is precisely equal to $n_\mathrm{tot}$; thus a reduction in the slope of $V_{\Delta n}$ as a function of $n_\mathrm{tot}$ from unity is associated with a quantum correlated signal and idler, which we refer to as photon number difference squeezing. In Fig. \ref{fig:tes_fig_exp}(b) the measured behaviour of $V_{\Delta n}$ as a function of $n_\mathrm{tot}$ is plotted for both coherent state inputs and for the generated signal and idler from the microring device. A reduction in the slope to ${\sim}0.70$ is observed, representing clearly evident photon number difference squeezing. At the highest powers (i.e., for measurements with the best signal-to-noise ratio) the ratio $V_{\Delta n}/n_\mathrm{tot}$ is suppressed to $0.704(5)$, representing $1.5(3)$~dB of number difference squeezing. This ratio, sometimes referred to as the ``noise reduction factor" \cite{harder2016single}, is a metric for nonclassicality \cite{aytur1990pulsed}, with a value below unity indicating a nonclassical state. For lower powers resulting in $n_\mathrm{tot}\ll 1$, the signal-to-noise ratio (likely limited by pump light leakage and broadband linear scattering processes) worsens and the squeezing degrades slightly (inset in Fig. \ref{fig:tes_fig_exp}(b)), though a significant sub-unity noise reduction factor is still observed even for the lowest powers used. This degradation effect at low powers was much more strongly pronounced in the samples with lower quality factor, consistent with our interpretation. The results exhibited in Fig. \ref{fig:tes_fig_exp} are therefore primarily limited by losses; as with the quadrature squeezing results, our directly measured system efficiency estimates are consistent with the amount of photon number squeezing observed. Unlike quadrature squeezing, in principle the magnitude of photon number difference squeezing is not gain limited: for an ideal lossless and noiseless SFWM process $V_{\Delta n}=0$ for all pump powers (assuming perfect detection efficiency), as signal and idler photons are always created in pairs. Thus the degree of number correlations on the chip is limited only by the finite escape efficiency of the resonator, and from this we infer that ${\sim}7$~dB of such correlations are available on-chip. This can be improved by designing a ring-channel coupler that facilitates higher escape efficiency, albeit at the expense of generation efficiency \cite{vernon2016no}.

A notable feature of this experiment is the high rate at which correlated multi-photon events are detected: at the maximum power used, coincidence events with ten detected photons were recorded at rates of $9.6(3)\times 10^2$ counts/s. In contrast, quantum photonic technologies based on chip-integrated parametric single photon sources must operate in the weakly driven regime, in which the probability of generating a photon pair is quite low, often well below $1$\%. As such, applications requiring multiple photons typically suffer from extremely low count rates. Our results demonstrate that many-photon states can be generated in a nanophotonic platform at much higher rates, motivating progress on applications that demand squeezed light sources, rather than single photons.

\bigskip
\subsection{Temporal mode structure}
\begin{figure}[h!]
\centering
\includegraphics[width=0.7\textwidth]{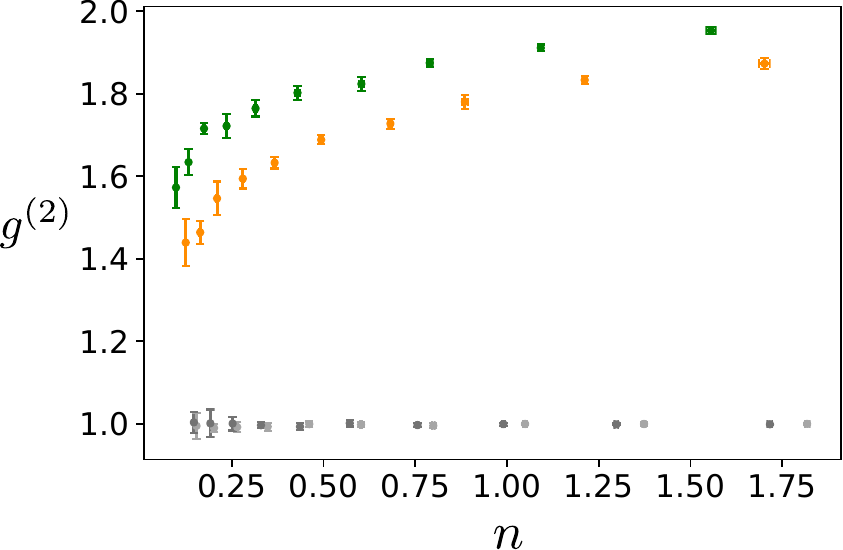}
\caption{Measured second-order correlation $g^{(2)}$ (Eq. (\ref{eq:g2})) for the signal and idler (green and orange points, respectively), and for coherent states (grey points) for different values of photon number $n$. Data is collected over a range of pump pulse energies corresponding approximately to $0.1$ nJ to $1$ nJ on-chip. Coherent state data was acquired using attenuated laser pulses, and shows excellent agreement with the predicted $g^{(2)}=1$ for coherent states.}\label{fig:g2plot}
\end{figure}

As discussed in the Introduction, another crucial feature of  squeezed light sources for quantum sampling applications -- and indeed most experiments utilizing photon counting on continuous variable states -- is the temporal mode structure of the generated squeezed states. Squeezed states in a single temporal mode must be sought in order to yield the correct desired statistics and enable sampling applications to generate outputs that are faithfully sampled from the desired probability distribution. Sources which generate squeezing in multiple modes that cannot be easily separated will yield statistics that do not reflect the desired single mode features.  A simple example of this is as follows: considered independently, the signal and idler modes in an ideal two-mode squeezed state are each individually described by a thermal state, and therefore should display an unheralded second-order correlation value of $g^{(2)}=2$. If, however, the state contains the product of many such two-mode squeezed states, each of which populates a different temporal mode, and the measurement of the state cannot distinguish between photons that were detected from different temporal modes, then the second-order correlation will exhibit $g^{(2)}<2$; in the limit of very many equally populated modes, the photon number statistics become Poissonian and $g^{(2)}\to 1$. In quantum sampling applications employing squeezed light, all relevant information is extracted directly from the photon statistics of the output \cite{huh2015boson,arrazola2018using,bradler2017gaussian,schuld2020measuring,banchi2019molecular}; thus using a squeezed light source with a highly multi-mode temporal structure would destroy the utility of such a machine. It is therefore important to assess the temporal mode structure of the generated squeezed states from our source \cite{christ2011probing}.

In microresonator systems, the temporal mode structure of generated parametric fluorescence can be tuned by driving the device with pump pulses having varying pulse duration \cite{helt2010spontaneous}. This is a feature common to many parametric fluorescence sources:  the use of broadband (i.e., temporally short) pump pulses, having less well-defined photon energies, leads to a weaker degree of spectral correlations between the generated signal and idler photons. Since the presence of spectral correlations is equivalent to the presence of multiple populated temporal modes in the generated beams, short pulses are therefore required to achieve single temporal mode operation. Here the relevant bandwidth scale is the parametric gain bandwidth, i.e., the resonator linewidth or inverse dwelling time.  Previously this has been experimentally explored primarily in the context of \textit{weakly} driven spontaneous four-wave mixing for photon pair generation \cite{silverstone2015qubit}. In that context, and also in our regime of more strongly driven fluorescence for squeezing, pumping with a pulse that is comparable to or shorter than the resonator dwelling time $\tau_\mathrm{dwell} = 2Q/\overline{\omega}$ will result in nearly single temporal mode operation. However, unlike in the weakly driven context, here the pump is sufficiently strong that self- and cross-phase modulation also play a role in modifying the temporal mode structure of the generated light. Detailed modelling and simulations to identify the optimal point in parameter space will be left for a future report; here we merely demonstrate that our device approaches the desired operational point with $g^{(2)}$ nearly $2$, without any special engineering of the device \cite{vernon2017truly} or intricate tailoring of the pump pulse shape \cite{christensen2018engineering} beyond simple tuning of the pulse duration.

To assess the temporal mode features of our source, we used the same measured photon number data as that used to extract the degree of photon number difference correlations, and instead calculated the independent second-order correlation statistic $g^{(2)}_{S(I)}$ for the signal (idler), defined as:
\begin{eqnarray}\label{eq:g2}
g^{(2)}_{S(I)} = \frac{\langle n_{S(I)}^2\rangle - \langle n_{S(I)}\rangle}{\langle n_{S(I)}\rangle^2}.
\end{eqnarray}
For two-mode squeezed vacuum states, this statistic provides a direct measure of the multi-mode character of the measured light; ideally, $g^{(2)}_{S(I)}=2$ for both the signal and idler, which would indicate perfect single temporal mode operation. The results are plotted in Fig. \ref{fig:g2plot} (green and orange points) for data collected with a range of pump pulse energies, leading to a range of generated photon numbers $n$ in the signal and idler; also plotted are the  measured $g^{(2)}$ values for attenuated laser pulses, i.e., coherent states (grey points). The pulses used had approximately rectangular intensity profile and duration 1.5 ns, which is about $1.2\times \tau_\mathrm{dwell}$.  The highest measured second-order correlations for the signal and idler are $g^{(2)}_S=1.95(1)$ and $g^{(2)}_I=1.87(1)$.  This raw measured $g^{(2)}$ for the signal is comparable to previous demonstrations using specially engineered waveguide sources \cite{eckstein2011highly,harder2013optimized,harder2016single}. The idler exhibited a consistently slightly lower second-order correlation value than that of the signal; we attribute this mismatch to an imbalance in the quantity of noise on each channel, and we estimate this noise to be between $0.02$ and $0.07$ measured photons per pulse for each channel, primarily from Raman scattering in the fiber components used. 

\bigskip
\section{DISCUSSION}\label{sec:discussion}
That our source can generate nearly single temporal squeezed states without special engineering of dispersion or phase matching is primarily due to its \textit{resonant} nature. Resonators provide an intrinsic narrow confinement of the nonlinear gain spectrum, owing to the narrow linewidths associated with their resonances, making it easier avoid populating unwanted temporal modes with photons. The measured $g^{(2)}$ quantities are not upper bounds: by selectively engineering the quality factors of the pump, signal, and idler resonances\cite{vernon2017truly}, or engineering the pump pulse more carefully \cite{christensen2018engineering}, squeezed states with $g^{(2)}$ arbitrarily close to 2 could be generated. 

As with most integrated quantum optical devices, losses are the primary factor that limit performance. In our case the ${\sim}1.5$~dB of coupling losses associated with extracting the squeezed light to a fiber limits the squeezing obtainable \textit{off} chip (assuming a new device with perfect ring escape efficiency, and assuming perfect collection and detection efficiencies) to about $5$~dB for both quadrature and photon number measurement contexts. Lowering the outcoupling loss to $0.5$ dB -- a challenging but not unrealistic goal -- would immediately improve this bound to nearly $10$ dB, at which point the resonator escape efficiency would become the dominant factor limiting the available squeezing.  Improving on-chip propagation loss would result in higher resonator quality factors, enabling higher escape efficiencies to be realized while maintaining or improving power efficiency, leading to better squeezing. We estimate that best-reported microring resonator devices are capable of generating over $10$~dB of on-chip quadrature squeezing with under $100$~mW of pump power \cite{vernon2019scalable}. Improving the generation efficiency would also improve the signal-to-noise ratio by reducing the amount of power needed to operate at the desired level of squeezing, thereby reducing the quantity of photons generated from spontaneous Raman scattering in fiber components. 

For some applications, including the most general CV quantum sampling architectures, it is convenient or necessary to generate squeezing in a single ``degenerate'' mode confined to a small frequency range, rather than distributed over a pair of signal and idler modes. This can be accomplished in our device by employing a dual pump scheme, which has already been used for degenerate photon pair sources \cite{he2014degenerate} and OPOs \cite{okawachi2015dual}. However, care must be taken with such a device to suppress spurious SFWM processes involving the squeezed resonance and unwanted resonances; without such a suppression scheme, an appreciable amount of excess noise is predicted to contaminate the output. This has been theoretically discussed \cite{vernon2019scalable}, and is a natural next step for engineering chip-integrated squeezed light sources \cite{zhang2020single,zhao2020near}.

\bigskip
\section{Materials and methods}
For all experiments light was coupled into the chip using ultrahigh numerical aperture (UHNA7) optical fibers manually aligned to the chip facets. Index matching gel was used to reduce reflections at the interface. The chip was temperature stabilized using a thermo-electric cooler with a slow PID feedback loop, and electrical probes were used to access the microheater elements. The radii of the microring resonators used were $120$~$\upmu\mathrm{m}$ and $200$~$\upmu\mathrm{m}$ for the quadrature squeezing and photon number difference squeezing experiments, respectively.

Quadrature squeezing measurements were performed using a commercial balanced receiver (Wieserlabs BPD1GA), which has approximately $80\%$ quantum efficiency at $1550$nm. Balancing was performed using a tunable fiber-coupled beam splitter (Newport). Linearity of the detector was assessed by varying the local oscillator level while monitoring the difference photocurrent fluctuations on an electrical spectrum analyzer; the local oscillator was then set to provide $13$dB of dark noise clearance, well into the linear regime but far short of saturation, at an optical power of $6$ mW. The quality factors of the resonances used in this experiment were approximately $2\times 10^5$, with escape efficiencies of approximately 75\%. The pump, signal, and idler wavelengths were approximately $1550.2$nm, $1548.7$nm, and $1551.7$nm, respectively. The free spectral range of the resonator was approximately $190$~GHz.  

Photon number resolving measurements were carried out using transition edge sensors, read out by inductive coupling to a cryogenic pre-amplifier based on low-noise coherent SQUID arrays; this system was provided by the National Institute for Standards and Technology. Traces from the detection system output were further amplified and then digitized by an analog-to-digital converter (AlazarTech ATS9440), which was triggered by an electronic pulse synchronized to the optical pulses. The traces were then analyzed in software to assign each one an integer corresponding to its most probable photon number, using a technique based on principal component analysis \cite{humphreys2015tomography}. We estimate the photon miscategorization probability to be on order $10^{-3}$ in the regime studied. Coherent state calibration data was obtained by heavily attenuating a pulsed laser source incident on the detectors and measuring the corresponding statistics to verify they conform to a Poissonian distribution that agrees with the mean photon number. To avoid saturating the detectors, the photon number experiments were carried out with a pump pulse train with $62.5$kHZ repetition rate. The quality factors of the resonances used in this experiment were approximately $8\times 10^5$, with escape efficiencies of approximately 80\%. The pump, signal, and idler wavelengths were approximately $1554.2$nm, $1551.4$nm, and $1557.0$nm, respectively. The free spectral range of the resonator was approximately $175$~GHz. 

Photon number data was taken in sets of 800,000 samples for each data point. Each set was used to develop a template for assigning integer photon numbers to the electrical pulses that are sampled from the transition edge sensor detectors. The resultant 800,000 integer photon numbers were divided into eight subsets of 100,000 samples. Quantities like noise reduction factor, number difference variance, and second-order correlation statistics were calculated for each of these subsets; the resultant means and standard deviations were used for the points and error bars on Figs. \ref{fig:tes_fig_exp} and \ref{fig:g2plot}.

Further detail on these and other aspects of the experimental apparatus and technique is available in the Supplementary Material. All data needed to evaluate the conclusions in the paper are present in the paper and/or the Supplementary Materials. Additional data available from authors upon request.

\bigskip
\subsection*{Acknowledgments}
We thank A. Dutt and R. Slavik for helpful discussions. 
Note regarding trade names: We use trade names to specify the experimental procedure adequately and do not imply endorsement by the National Institute of Standards and Technology.  Similar products or services provided by other manufacturers or vendors may work as well or better. Competing Interests: The authors declare that they have no competing interests. Author contributions: VDV performed the quadrature experiments. BM participated in building the pump systems for both the quadrature and photon number experiments, supervised chip device design, and wrote part of the supplement. LGH performed theoretical modelling, developed the TES pulse discrimination code, and wrote part of the supplement. NQ performed theoretical modelling. RS, DHM, and MJC performed the photon counting experiments. KT and MM participated in chip device design. JL, AR, and RCP provided guidance on selecting measurements to present. AEL, TG, and SWN constructed the TES system. ZV supervised the project and wrote the manuscript.

\bigskip
\bibliographystyle{Science}
\bibliography{references}

\newpage
\setcounter{equation}{0}
\setcounter{figure}{0}
\setcounter{table}{0}
\makeatletter
\renewcommand{\theequation}{S\arabic{equation}}
\renewcommand{\thefigure}{S\arabic{figure}}
\renewcommand{\thesection}{S\arabic{section}}
\begin{center}\section*{Supplementary Materials}\end{center}

\section*{S1. Quadrature variance measurement}
\subsection*{Detailed Experimental Apparatus}
The quadrature variance measurement (Fig.~2(a) of main text) is performed using three resonances of the microring that span a bandwidth of $3$~nm. The central resonance of the microring is pumped by a fiber-amplified external cavity diode laser (ECDL). Amplified spontaneous emission generated near the signal and idler frequencies is filtered out by a WDM filter placed at the input of the chip. A polarization controller on the pump fiber aligns the input polarization to the TE mode of the on-chip waveguides.

The frequencies of the microring resonances can be tuned using an integrated microheater, and are actively stabilized to the pump wavelength through a side-of-peak lock. The transmitted pump beam is separated from the generated signal and idler wavelengths using an (off-chip) WDM filter, and its extinction is monitored on a photodetector. This signal is sent through a PID filter and fed back to the integrated heater to stabilize the microring resonances to the pump frequency. 

\begin{figure}[ht]
\centering
\includegraphics{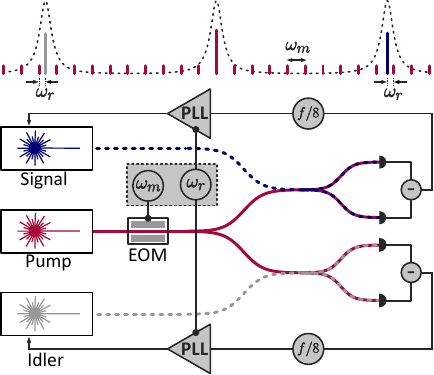}
\caption{Phase-locking schematic for the bichromatic LO. The pump is strongly phase-modulated at $\omega_m$ to produce sidebands near the signal and idler resonances. The reference clock $\omega_r$ used to stabilize the signal and idler beatnotes is derived from same source as $\omega_m$ to minimize excess phase-noise. EOM: electro-optic modulator.}
\label{fig:PLL}
\end{figure}

\begin{figure}[ht]
\centering
\includegraphics{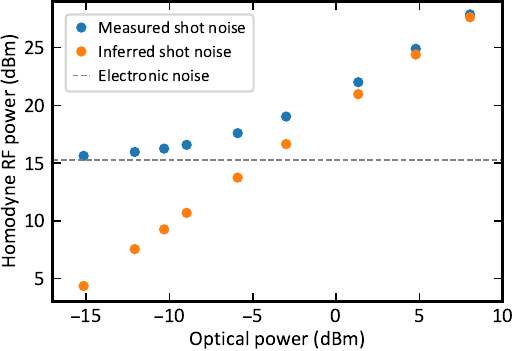}
\caption{Shot noise calibration of the bichromatic LO. The noise integrated over the $1$~GHz bandwidth of the homodyne detector is measured as a function of the total optical power in both LO beams. All measurements presented in the main text are performed with an LO optical power of $8$~dBm, well above the detector noise floor.}
\label{fig:shotcal}
\end{figure}

The filtered signal and idler photons are carried over the same optical fiber, to a tunable beamsplitter where they are interfered with the bichromatic local oscillator (LO) for balanced homodyne detection. The bichromatic LO is generated by combining light from two ECDLs operating at the signal and idler wavelengths as shown in Fig.~\ref{fig:PLL}. Variable attenuators and polarization controllers placed inline with each LO provide independent power and polarization control of the two beams at the balanced homodyne detector. The power in each LO beam is independently adjusted to provide $10$~dB of shot-noise clearance above the electronic noise floor of the homodyne detector. The shot-noise of the resulting bichromatic LO is thus $13$~dB above the detector noise floor (see Fig.~\ref{fig:shotcal}). Their polarizations are also independently adjusted to align with those of the generated signal and idler at the detector.

The signal and idler ECDLs are stabilized to the pump phase as illustrated in Fig.~\ref{fig:PLL}. Low power pickoffs from the pump, signal and idler LO lasers (see Fig 1) are used to measure the relative phase between pump and LO lasers. A strong phase modulation at $\omega_m = 2\pi \times 15.694$~GHz is applied to the pump to generate sidebands $188$~GHz away, near the signal and idler frequencies. The signal and idler lasers are interfered with the modulated pump beam to produce RF beatnote signals at $800$~MHz on a pair of homodyne detectors.

A pair of phase-locked loop (PLL) controllers stabilize the phases of these beatnotes to a $\omega_r=2\pi \times 100$~MHz reference clock by feeding back on fast current modulation ports present on both the signal and idler ECDLs. The reference clock is derived from the same synthesizer that modulates the pump, to avoid adding excess phase noise to the PLL. Our feedback scheme suppresses relative phase noise between the pump and each LO beam to within $4^{\circ}$~RMS and ensures that the signal and idler LOs are symmetrically split about the pump frequency as illustrated in Fig.~\ref{fig:PLL}.

The squeezing measurement presented in Fig.~2 of the main text is performed by analyzing the homodyne detector output on an RF spectrum analyzer, as the phase of the bichromatic LO is varied using a fiber-strecher. The noise level at $20$~MHz is measured with a resolution bandwidth (RBW) of $1$~MHz and a video bandwidth (VBW) of $300$~Hz, while the LO phase is ramped at an approximate rate of $2\pi$~s$^{-1}$. The squeezing spectrum measurements presented in Fig.~3 of the main text are performed under similar conditions, while recoding the maximum and minumum values of the noise level at each sampled frequency.

\begin{figure*}[t]
\centering
\includegraphics[width=\textwidth]{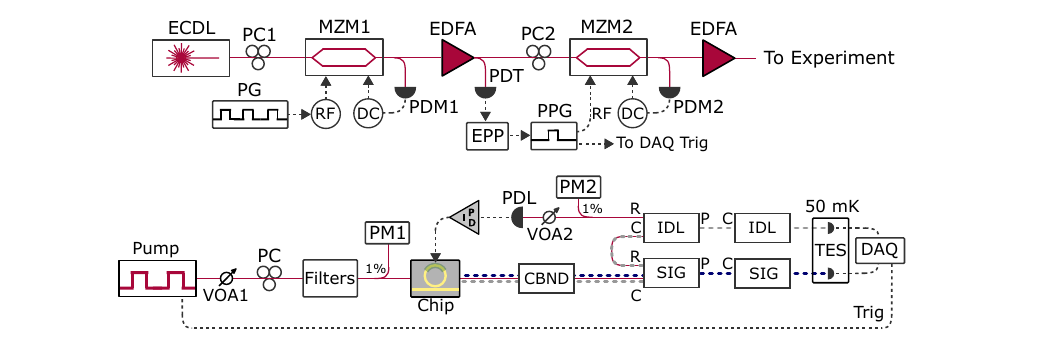}
\caption{Detailed experimental setup of the photon number measurements. Further description is found in the text.}
\label{fig:SupPNRExp}
\end{figure*}

\subsection*{Theory}
Equations to which the data is fit are derived as follows. Adding the appropriate channel-ring coupling Hamiltonian~(\textit{27}) to the nonlinear Hamiltonian of Eq.~(1) in the main text, the Fourier transformed equations of motion for the signal and idler ring mode operators can be written
\begin{align}
-i\Omega b_{\mathrm{S}}\left(\Omega\right)&=-\tilde{\Gamma}_{\mathrm{S}}b_{\mathrm{S}}\left(\Omega\right)-i\gamma_{\mathrm{S}}^{*}\psi_{\mathrm{S}_{\text{in}}}\left(\Omega\right)\nonumber\\
&\quad-i\mu_{\mathrm{S}}^{*}\phi_{\mathrm{S}_{\text{in}}}\left(\Omega\right)+i\Lambda\bar{\beta}_{\mathrm{P}}^{2}b_{\mathrm{I}}^{\dagger}\left(-\Omega\right),\nonumber\\
-i\Omega b_{\mathrm{I}}^{\dagger}\left(-\Omega\right)&=-\tilde{\Gamma}_{\mathrm{I}}^{*}b_{\mathrm{I}}^{\dagger}\left(-\Omega\right)+i\gamma_{\mathrm{I}}\psi_{\mathrm{I}_{\text{in}}}^{\dagger}\left(-\Omega\right)\nonumber\\
&\quad+i\mu_{\mathrm{I}}\phi_{\mathrm{I}_{\text{in}}}^{\dagger}\left(-\Omega\right)-i\Lambda\left(\bar{\beta}_{\mathrm{P}}^{*}\right)^{2}b_{\mathrm{S}}\left(\Omega\right),
\label{eq:coupled}
\end{align}
where $\psi_{x_{\text{in}}}$ are channel mode operators with input-output relations
\begin{equation}
\psi_{x_{\text{out}}}\left(\Omega\right)=\psi_{x_{\text{in}}}\left(\Omega\right)-\frac{i \gamma_{x}}{v_{x}}b_{x}\left(\Omega\right),
\label{eq:input-output}
\end{equation}
and $\phi_{x_{\text{in}}}$ are ``phantom channel'' mode operators that allow one to include additional sources of loss easily. Here
$\left\vert\gamma_{x}\right\vert^{2}+\left\vert\mu_{x}\right\vert^{2}=2 v_{x}\bar{\Gamma}_{x}$, $\left\vert\gamma_{x}\right\vert^{2}=2 v_{x}\bar{\Gamma}_{x}\eta_{x}$, and $\tilde{\Gamma}_{x}=\bar{\Gamma}_{x}-i \Delta$, where $\Delta$ is an effective detuning including pump detuning from resonance as well as the power-dependent effects of self- and cross-phase modulation. The channel operators satisfy the commutation relations
\begin{align}
\left[\psi_{x_{\text{in}}}\left(\Omega\right),\psi_{y_{\text{in}}}^{\dagger}\left(\Omega^{\prime}\right)\right]&=\left[\phi_{x_{\text{in}}}\left(\Omega\right),\phi_{y_{\text{in}}}^{\dagger}\left(\Omega^{\prime}\right)\right]\nonumber\\
&=\delta_{xy}\delta\left(\Omega-\Omega^{\prime}\right)/v_{x}.
\end{align}

Equations~\eqref{eq:coupled} are readily solved as 
\begin{align}
b_{\mathrm{S}}\left(\Omega\right)&=\frac{\bar{\beta}_{\mathrm{P}}^{2}\Lambda\left[\gamma_{\mathrm{I}}\psi_{\mathrm{I}_{\text{in}}}^{\dagger}\left(-\Omega\right)+\mu_{\mathrm{I}}\phi_{\mathrm{I}_{\text{in}}}^{\dagger}\left(-\Omega\right)\right]}{\left|\bar{\beta}_{\mathrm{P}}\right|^{4}\Lambda^{2}-\left(\tilde{\Gamma}_{\mathrm{I}}^{*}-i\Omega\right)\left(\tilde{\Gamma}_{\mathrm{S}}-i\Omega\right)}\nonumber\\
&\quad+\frac{i\left[\gamma_{\mathrm{S}}^{*}\psi_{\mathrm{S}_{\text{in}}}\left(\Omega\right)+\mu_{\mathrm{S}}^{*}\phi_{\mathrm{S}_{\text{in}}}\left(\Omega\right)\right]\left(\tilde{\Gamma}_{\mathrm{I}}^{*}-i\Omega\right)}{\left|\bar{\beta}_{\mathrm{P}}\right|^{4}\Lambda^{2}-\left(\tilde{\Gamma}_{I}^{*}-i\Omega\right)\left(\tilde{\Gamma}_{\mathrm{S}}-i\Omega\right)},\nonumber\\
b_{\mathrm{I}}^{\dagger}\left(-\Omega\right)&=\frac{\left(\bar{\beta}_{\mathrm{P}}^{*}\right)^{2}\Lambda\left[\gamma_{\mathrm{S}}^{*}\psi_{\mathrm{S}_{\text{in}}}\left(\Omega\right)+\mu_{\mathrm{S}}^{*}\phi_{\mathrm{S}_{\text{in}}}\left(\Omega\right)\right]}{\left|\bar{\beta}_{\mathrm{P}}\right|^{4}\Lambda^{2}-\left(\tilde{\Gamma}_{\mathrm{I}}^{*}-i\Omega\right)\left(\tilde{\Gamma}_{\mathrm{S}}-i\Omega\right)}\nonumber\\
&\quad\frac{-i\left[\gamma_{\mathrm{I}}\psi_{\mathrm{I}_{\text{in}}}^{\dagger}\left(-\Omega\right)+\mu_{\mathrm{I}}\phi_{\mathrm{I}_{\text{in}}}^{\dagger}\left(-\Omega\right)\right]\left(\tilde{\Gamma}_{\mathrm{S}}-i\Omega\right)}{\left|\bar{\beta}_{\mathrm{P}}\right|^{4}\Lambda^{2}-\left(\tilde{\Gamma}_{\mathrm{I}}^{*}-i\Omega\right)\left(\tilde{\Gamma}_{\mathrm{S}}-i\Omega\right)},
\end{align}
and can then be used, along with Eqs.~(\ref{eq:coupled}), to solve for the moments
\begin{align}
N_{x}\left(\Omega,\Omega^{\prime}\right)\delta\left(\Omega-\Omega^{\prime}\right)&=v_{x}\left\langle \psi_{x_{\text{out}}}^{\dagger}\left(\Omega\right)\psi_{x_{\text{out}}}\left(\Omega^{\prime}\right)\right\rangle,\nonumber\\
M_{xy}\left(\Omega,\Omega^{\prime}\right)\delta\left(\Omega+\Omega^{\prime}\right)&=\sqrt{v_{x}v_{y}}\left\langle\psi_{x_{\text{out}}}\left(\Omega\right)\psi_{y_{\text{out}}}\left(\Omega^{\prime}\right)\right\rangle,
\end{align}
or
\begin{align}
N_{x}\left(\Omega,\Omega\right)&=\frac{4\eta_{x}\bar{\Gamma}_{\mathrm{S}}\bar{\Gamma}_{I}\left|\bar{\beta}_{\mathrm{P}}\right|^{4}\Lambda^{2}}{\left|\left|\bar{\beta}_{\mathrm{P}}\right|^{4}\Lambda^{2}-\left(\tilde{\Gamma}_{\mathrm{I}}^{*}-i\Omega\right)\left(\tilde{\Gamma}_{\mathrm{S}}-i\Omega\right)\right|^{2}}\;\;,\nonumber\\
M_{xy}\left(\Omega,-\Omega\right)&=2\sqrt{\eta_{\mathrm{S}}\bar{\Gamma}_{\mathrm{S}}\eta_{I}\bar{\Gamma}_{I}}\left|\bar{\beta}_{\mathrm{P}}\right|^{2}\Lambda\nonumber\\
&\quad\times\frac{\left|\bar{\beta}_{\mathrm{P}}\right|^{4}\Lambda^{2}+\left(\tilde{\Gamma}_{x}^{*}+i\Omega\right)\left(\tilde{\Gamma}_{y}^{*}-i\Omega\right)}{\left|\left|\bar{\beta}_{\mathrm{P}}\right|^{4}\Lambda^{2}-\left(\tilde{\Gamma}_{\mathrm{I}}^{*}-i\Omega\right)\left(\tilde{\Gamma}_{\mathrm{S}}-i\Omega\right)\right|^{2}}\;\;,
\end{align}
up to an inconsequential global phase on $M_{xy}\left(\Omega,-\Omega\right)$. Note that $N_{x}\left(\Omega,\Omega\right)^{*}=N_{x}\left(\Omega,\Omega\right)$ and $M_{xy}\left(\Omega,-\Omega\right)=M_{yx}\left(-\Omega,\Omega\right)$. Finally, we solve for the general quadrature variance at sideband frequency $\Omega$ relative to that where the signal is maximal with a bichromatic LO
\begin{align}
V\left(\phi_{\mathrm{S}},\phi_{I};\Omega\right)&=1+\bar{N}\left(\Omega,\Omega\right)+\bar{N}\left(-\Omega,-\Omega\right)\nonumber\\
&\quad+2\Re\left\{ e^{-i\left(\phi_{\mathrm{S}}+\phi_{I}\right)}\bar{M}\left(\Omega,-\Omega\right)\right\},
\end{align}
where the $\phi_{x}$ are local oscillator phases, $\bar{N}\left(\Omega,\Omega^{\prime}\right)=\left[N_{\mathrm{S}}\left(\Omega,\Omega^{\prime}\right)+N_{\mathrm{I}}\left(\Omega,\Omega^{\prime}\right)\right]/2$, and $\bar{M}\left(\Omega,\Omega^{\prime}\right)=\left[M_{\mathrm{SI}}\left(\Omega,\Omega^{\prime}\right)+M_{\mathrm{IS}}\left(\Omega,\Omega^{\prime}\right)\right]/2$. It has a form similar to that using a single local oscillator, as expected from bichromatic LO theory~(\textit{25}). For $\bar{\Gamma}_{\mathrm{S}}=\bar{\Gamma}_{I}\equiv\bar{\Gamma}$, $\eta_{\mathrm{S}}=\eta_{I}\equiv\eta$, and $\tilde{\Gamma}_{\mathrm{S}}=\tilde{\Gamma}_{I}\equiv\tilde{\Gamma}$, the variance can be written as
\begin{align}
&V\left(\phi_{\mathrm{S}},\phi_{I};\Omega\right)\nonumber\\
&=1+\frac{4\eta g\left(2g+\Re\left\{ e^{-i\left(\phi_{\mathrm{S}}+\phi_{I}\right)}\left[\left(\frac{\tilde{\Gamma}^{*}}{\bar{\Gamma}}\right)^{2}+g^{2}+\left(\frac{\Omega}{\bar{\Gamma}}\right)^{2}\right]\right\} \right)}{\left[g^{2}-\left(\frac{\left|\tilde{\Gamma}\right|}{\bar{\Gamma}}\right)^{2}+\left(\frac{\Omega}{\bar{\Gamma}}\right)^{2}\right]^{2}+4\left(\frac{\Omega}{\bar{\Gamma}}\right)^{2}},
\end{align}
where we have put $\left|\bar{\beta}_{\mathrm{P}}\right|^{2}\Lambda/\bar{\Gamma}=g$.

We note that our derivation parallels previous work on squeezing in $\chi^{\left(2\right)}$ optical parametric oscillators~(\textit{44-46}), however there are two important differences. The first is that because we consider a $\chi^{\left(3\right)}$ process, $g$ depends linearly on the number of pump photons (or, equivalently, the pump power) rather than its square root. The second is that a common consideration, when there are no power-dependent terms in $\Delta$, is to work such that $\Delta=0$ and thus, maximizing/minimizing over $\phi_{\mathrm{S}}+\phi_{I}$, denoted with the superscript $\pm$,
\begin{equation}
V^{\pm}\left(\Omega\right)_{\Delta=0}=1\pm\frac{4 g\eta}{\left(1\mp g\right)^{2}+\left(\frac{\Omega}{\bar{\Gamma}}\right)^{2}}.
\label{eq:Delta0}
\end{equation}
Yet here, we track the shifts due to self- and cross-phase modulation and work such that $\Delta=g\bar{\Gamma}
 $, leading to
\begin{equation}
V^{\pm}\left(0\right)_{\Delta=g\bar{\Gamma}}=1+4g\eta\left(2g\pm\sqrt{1+4g^{2}}\right),
\end{equation}
exactly as in Eq.~(2) of the main text. Note that here, unlike in \eqref{eq:Delta0}, $g$ is not constrained to be $\le 1$.

\section*{S2. Photon Number Measurement}
\subsection*{Detailed Experimental Apparatus}
A detailed schematic of the setup used in the photon number experiments is shown in Fig.~\ref{fig:SupPNRExp}. The setup comprises of four main parts: the pump source preparation, chip input filtering, the photonic chip coupling and control, and the post chip filtering and data acquisition. We will elaborate on the construction of these parts further below.
 
The pump section is comprised of two stages of Mach-Zehnder Modulators (MZM) and amplification, the first stage creates a high repetition pulse train. A continuous wave external cavity laser (Toptica CTL) sent to a \SI{20}{\GHz} Mach-Zehnder Modulator (MZM1, Optilab IM-1550-20). The DC setpoint of the modulator is set to null transmission in a closed feedback loop through the use of a homebuilt system using an FPGA to perform a dither lock (Red Pitaya 125-14), with a \SI{1}{\percent} tap of the output of the Modulator being measured at a photodetector as the input signal (PD1). An electrical pulse train drives the RF input of the Modulator (PG), with a pulse width of \SI{1.5}{\ns} and a repetition rate of \SI{64}{MHz} being generated by an arbitrary waveform generator (Agilent M8195A) and further amplified with a broadband RF amplifier (iXblue DR-PL-20-MO). The pulse train has a stable extinction ratio above \SI{30}{\dB}, attained through the use of the DC stabilization and alignment of the input polarization state with PC1. The pulse train is then amplified using a pulsed amplifier (Pritel FA-23-IO) to an average power of a few 10s of \si{\mW}. 

The second stage of the pump is used as a pulse picker for the input pulse train, reducing the duty cycle to be compatible with the photon number resolving detectors. To do this we generate a trigger pulse for the pulse picking generator (PPG, Siglent SDG 6052X), through the use of an input optical pulse and a home-built electrical pulse picking circuit (EPP). The optical pulse is measured on a TTL photodector (PDT, Koheron PD200T) giving a constant peak voltage output, and is compared to a reference level on the EPP circuit. After detecting an initial optical pulse the EPP generates a trigger pulse, and then counts 1024 pulses ($2^{10}$) before resetting. Upon receiving a trigger the PPG generates a square pulse of \SI{8}{\ns} width, which drives the RF input of a high extinction MZM (MZM2, iXblue MXER-LN-20) to pick the optical pulse train, with a second synchronized output being used to generate a trigger pulse for the data acquisition. The picked electrical pulse is aligned to the optical pulse train by tuning the relative electrical delay upon receiving a trigger with the PPG. A slight jitter of less than \SI{1}{\ns} is observed, which is well below the width of the pulse picking pulse guaranteeing that only a single optical pulse is within each pulse picking window. The DC bias of the modulator is controlled with a second stabilization system in the same way as MZM1. The pulse train, now comprising of \SI{1.5}{\ns} square pulses at a rep rate of \SI{62.5}{KHz}, is finally amplified by a second EDFA before being sent to the experiment. During operation the modulation bias control and pulse picking is monitored and the system was found to be stable over the course of a day.

After the pump source the input power to the chip is controlled with a computer controlled variable optical attenuator (VOA1) with a fixed attenuation uncertainty of \SI{0.1}{\dB} and attenuation precision of \SI{0.001}{\dB} (VIAVI mVOA-C1). A bank of filters are used to remove unwanted residual photons from the pump light. The filters consisted of the following: two \SI{980}{\nm} filters (Thorlabs WD9850AB), required due to the pumping diodes in the EDFAs, followed by two \SI{100}{\GHz} WDMs centered around the pump at \SI{1549.9}{\nm} (Opneti \SI{100}{\GHz} DWDM), which reject amplified spontaneous emission (ASE) from the amplifiers, as well as spontaneous Raman scattering generated in the fiber patch cords in the setup. Each WDM provides greater than \SI{80}{\dB} rejection close to the passband, with a typical insertion loss of \SI{0.5}{\dB}. The fiber pigtail output of the final WDM was spliced down to minimize generated Raman being injected into the chip. 

The filtered pump pulses were coupled to the chip using small-core fiber (Nufern UHNA7) and index matching gel, mounted on multi-axis stages (Newport 562F) equipped with stepper motor actuators with \SI{25}{\nm} resolution (Zaber LAC10A). The insertion loss was monitored using optical power meters (Thorlabs PM100USB) located before and after the filtering stages (PM1, PM2), with the chip coupling bypassed to provide a reference measurement. During the experiment the ring was actively side-of-peak locked to the laser, through the use of a feedback signal generated from an FPGA board (PID, Red Pitaya 125-14 running PyRPL 0.9.4 (\textit{47})). When the pump power was swept for the experiment, using VOA1, the input power to the locking detector (PDL) was controlled with VOA2 to prevent saturation of the FPGA analog-to-digital converter. To ensure consistent behaviour for the measurements, the setpoint of the side-of-peak lock was calibrated for each separate pump power. The input polarization to the chip was optimized to maximize the extinction of the pump resonance of interest, using the input PC.

To ensure that only the desired generated light is measured on the TES detectors a series of filters wavelength division multiplexers (WDM) are used after the chip to reject the pump and direct the two desired frequency channels to separate detectors. Each WDM consists of an input ``common" port (C), the transmitted passband (P) and a reflection port (R). Immediately after the chip a bandpass filter (CBND) is used to only transmit light from \SI{1540}{nm} to \SI{1560}{nm}, filtering light down to \SI{1200}{nm}. As shown in Fig.~\ref{fig:SupPNRExp}, the reflected light from the initial WDM, aligned to the signal frequency (SIG), is input to the common port of another WDM which is aligned to the idler frequency (IDL). The reflected light from this WDM contains the pump frequency, which is used in the locking setup described above. A second WDM at both the idler and signal frequencies is used to ensure the pump is sufficiently rejected in the bands of interest. The \SI{117}{\GHz} free spectral range of the micro-ring resonances ensures that only a single resonance is within the passband of the \SI{100}{\GHz} WDMs. The loss of the two paths of the signal and idler was approximately \SI{1}{dB}.

Finally, the temporal traces of the TES detectors were acquired using a DAQ card (AlazarTech ATS9440), which was triggered (TRIG) with the same pulse generator used for generating the picked pump pulses (PPG). The temporal window of \SI{5}{\micro \second} contained a single pulse response from the detectors. In the next section we detail the processing of the acquired time traces to determine the photon number of individual measurements.

\subsection*{Data Acquisition and Processing}
\begin{figure}[ht]
\centering
\includegraphics[width=0.6\textwidth]{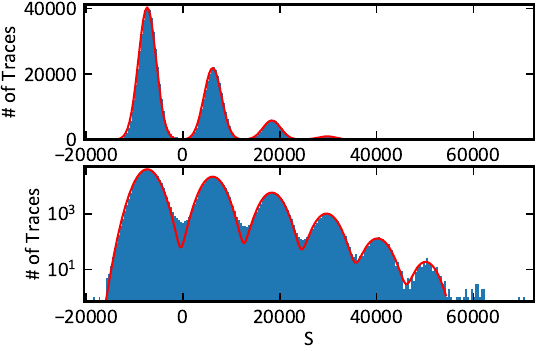}
\caption{Representative histogram of traces ordered according to overlap with their first principle component $s_{i}$ on both linear (top) and log (bottom) scales. The red line corresponds to a fit using a sum of 6 Gaussians.}
\label{fig:RepHist}
\end{figure}

Our data consists of sets of 800000 voltage traces $\bm{V}=\left\{v_{i}\left(t\right)\right\}$ in each of two channels, where each set corresponds to a different controlled amount of attenuation between source and detector. To assign a photon number to an individual trace in each set, first we follow Ref.~(\textit{43}) and perform a principal component analysis on each set $\bm{V}$, which amounts to solving for the eigenvectors of $\tilde{\bm{V}}^{T}\tilde{\bm{V}}$, where $\tilde{\bm{V}}$ is the data set with its mean trace subtracted. Calling these mean-subtracted traces $\tilde{v}_{i}\left(t\right)$ and their respective first principle component $PC\left(t\right)$, this enables ordering of the $v_{i}\left(t\right)$ within each set according to $\int\tilde{v}_{i}\left(t\right)PC\left(t\right)\text{d}t\equiv s_{i}$. We then construct a histogram of the $s_{i}$ with 223(${\sim} \sqrt{800000}/4)$ bins, and observe that they naturally cluster (see Fig.~\ref{fig:RepHist}). We associate these clusters with discrete photon numbers, and determine the $s_{i}\in S$ boundaries for discretization by fitting a sum of Gaussians to the histogram and solving for their points of intersection. Having ``digitized'' the $v_{i}\left(t\right)$ for each channel, it is then a simple matter to determine the mean photon number of the sum of the two channels $n_{\text{tot}}=\left\langle N_{\mathrm{S}}+N_{I}\right\rangle$ as well as the variance of the photon number difference of the two channels $V_{\Delta n}=V_{N_{\mathrm{S}} - N_{I}}$. We estimate that the largest source of error associated with this procedure arises from traces with an $s_{i}$ much larger than the Gaussian centered at the largest $s_{i}$, as all of these traces will be assigned photon number $n$ when they more likely correspond to $n+1$ or even $n+2$. However, we calculate that they form less than 0.04\% of the traces in any given set, and therefore lead to error bars on data presented in Fig.~4 of the main text too small to be seen by eye.

Consideration of other sources of error in this type of procedure is a difficult and relatively new problem, prime for future study. In the present work we consider each extracted integer sample as a separate, uncorrelated measurement. We estimate the \textit{statistical} uncertainty in each aggregate quantity -- $V_{\Delta n}$, $n_\mathrm{tot}$, and $g^{(2)}$ --  by calculating the corresponding quantity on eight 100,000-sample subsets of the 800,000 samples collected for each pump setting. The mean and standard deviation of these are used for the data points in Figs. 4 and 5 in the main text. This accurately quantifies the degree to which 100,000-sample measurements of such first- and second-order photon number moments vary between repeated independent measurements. Systematic uncertainties in these aggregate quantities can be estimated from the measurements on coherent states, where the corresponding quantities are extremely well modeled by Poisson statistics. The deviation of associated means from the predicted values gives an estimate of the relative systematic uncertainty in reported quantities, which is on the order of $10^{-3}$ for the NRF and $10^{-2}$ for $g^{(2)}$. 

\begin{figure}[ht]
\centering
\includegraphics[width=0.6\textwidth]{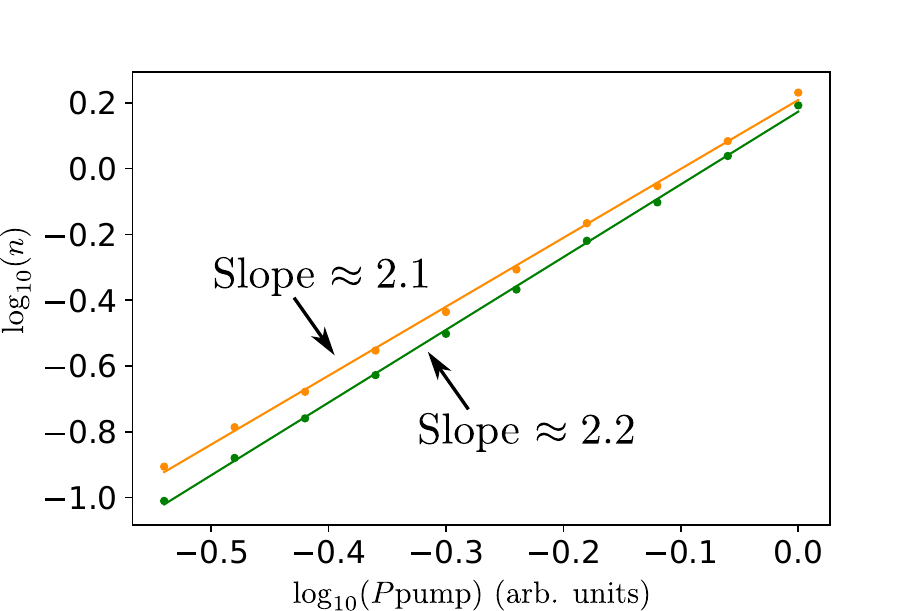}
\caption{Measured logarithm of $n_\mathrm{tot}$ vs. logarithm of injected pump power for values of $n_\mathrm{tot}>1$.}
\label{fig:loglog}
\end{figure}

Finally, we explore the predicted quadratic scaling of $n_s$ and $n_i$ with pump power for spontaneous four-wave mixing (SFWM) in a ring resonator when the pump detuning is locked to a ring resonance that moves due to self- and cross-phase modulation~(\textit{27}). In Fig.~\ref{fig:loglog} we plot the measured logarithm of $n_\mathrm{tot}$ as a function of the logarithm of injected pump pulse peak power to the ring $P_\mathrm{P}$ over the range of pump powers used, and observe that it is indeed nearly quadratic, with a slope close to 2 for both the signal and idler. 

\subsection*{Theory}
To calculate the expected slope of $V_{\Delta n}$ vs.\ $n_{\text{tot}}$, we first write the state generated on chip as the two-mode squeezed vacuum
\begin{equation}
\left|\psi\right\rangle =e^{\sum_{\ell}r_{\ell}a_{\mathrm{S}\ell}^{\dagger}a_{\mathrm{I}\ell}^{\dagger}-\text{H.c.}}\left|0\right\rangle _{\mathrm{S}}\left|0\right\rangle _{\mathrm{I}},
\end{equation}
for which the number operator in either the signal or idler channel is
\begin{equation}
N_{x}=\sum_{\ell}a_{x\ell}^{\dagger}a_{x\ell}.
\label{eq:nx}
\end{equation}
Introducing loss with transmission factors $\eta_{x}$ we have
\begin{align}
\left\langle a_{x\ell}^{\dagger}a_{y\ell^{\prime}}\right\rangle &=\delta_{xy}\delta_{\ell\ell^{\prime}}\eta_{x}\sinh^{2}r_{\ell},\nonumber\\
\left\langle a_{x\ell}a_{y\ell^{\prime}}\right\rangle &=\left(1-\delta_{xy}\right)\delta_{\ell\ell^{\prime}}\sqrt{\eta_{x}\eta_{y}}\sinh r_{\ell}\cosh r_{\ell}.
\end{align}
We can thus solve for the photon number difference variance
\begin{align}
V_{\Delta n} &=V_{N_{\mathrm{S}}}+V_{N_{\mathrm{I}}}-2\left(\left\langle N_{\mathrm{S}}N_{\mathrm{I}}\right\rangle -\left\langle N_{\mathrm{S}}\right\rangle \left\langle N_{\mathrm{I}}\right\rangle \right)\nonumber\\
&=\eta_{\mathrm{S}}\sum_{\ell}\sinh^{2}r_{\ell}\left(\eta_{\mathrm{S}}\sinh^{2}r_{\ell}+1\right)\nonumber\\
&\quad+\eta_{\mathrm{I}}\sum_{\ell}\sinh^{2}r_{\ell}\left(\eta_{\mathrm{I}}\sinh^{2}r_{\ell}+1\right)\nonumber\\
&\quad-2\eta_{\mathrm{S}}\eta_{\mathrm{I}}\sum_{\ell}\sinh^{2}r_{\ell}\cosh^{2}r_{\ell},
\end{align}
or, when $\eta_{\mathrm{S}}=\eta_{\mathrm{I}}\equiv\eta$,
\begin{equation}
V_{\Delta n}	=2\eta\left(1-\eta\right)\sum_{\ell}\sinh^{2}r_{\ell}=\left(1-\eta\right)n_{\text{tot}},
\end{equation}
exactly as in Eq.~(3) of the main text.

\end{document}